\documentclass[12pt]{article}

\usepackage{lmodern} 
\usepackage{amsmath,amsfonts,amssymb}
\usepackage{amsthm}   
\usepackage{latexsym}
\usepackage{natbib} 
\usepackage[nottoc]{tocbibind} 
\usepackage{algorithm2e}
\usepackage{microtype}
\usepackage{subcaption} 
\usepackage{multirow} 
\usepackage{booktabs} 
\usepackage{url}
\usepackage{afterpage,float,caption} 
\usepackage{rotating}   
\usepackage{graphicx}  
\usepackage{float}   
\usepackage{subcaption}
\usepackage{tablefootnote}
\usepackage{booktabs, caption, makecell} 
\usepackage{threeparttable} 

\makeatletter
\newcommand{\distas}[1]{\mathbin{\overset{#1}{\kern\z@\sim}}}%
\newsavebox{\mybox}\newsavebox{\mysim}
\newcommand{\distras}[1]{%
	\savebox{\mybox}{\hbox{\kern3pt$\scriptstyle#1$\kern3pt}}%
	\savebox{\mysim}{\hbox{$\sim$}}%
	\mathbin{\overset{#1}{\kern\z@\resizebox{\wd\mybox}{\ht\mysim}{$\sim$}}}%
}
\makeatother

\usepackage[T1]{fontenc}
\usepackage[utf8]{inputenc}
\usepackage[affil-it]{authblk}

\begin{document}

\title{\textbf{On the ``Poisson Trick'' and its Extensions for Fitting Multinomial Regression Models}}
\author[1 2]{Jarod Y.L. Lee}
\author[1 3]{Peter J. Green}
\author[1 2 4]{Louise M. Ryan}
\affil[1]{School of Mathematical and Physical Sciences, University of Technology Sydney, Australia.}
\affil[2]{Australian Research Council Centre of Excellence for Mathematical \& Statistical Frontiers, The University of Melbourne, Australia.}
\affil[3]{School of Mathematics, University of Bristol, U.K.}
\affil[4]{Department of Biostatistics, Harvard T.H. Chan School of Public Health, U.S.}
\date{\today}
\maketitle


\begin{abstract}
	This article is concerned with the fitting of multinomial regression models using the so-called ``Poisson Trick''. The work is motivated by \cite{Chen2001anote} and \cite{Malchow-Moller2003estimation} which have been criticized for being computationally inefficient and sometimes producing nonsense results. We first discuss the case of independent data and offer a parsimonious fitting strategy when all covariates are categorical. We then propose a new approach for modelling correlated responses based on an extension of the Gamma-Poisson model, where the likelihood can be expressed in closed-form. The parameters are estimated via an Expectation/Conditional Maximization (ECM) algorithm, which can be implemented using functions for fitting generalized linear models readily available in standard statistical software packages. Compared to existing methods, our approach avoids the need to approximate the intractable integrals and thus the inference is exact with respect to the approximating Gamma-Poisson model. The proposed method is illustrated via a reanalysis of the yogurt data discussed by \cite{Chen2001anote}. 
	

\end{abstract} 

\section{Introduction}
\label{sec:intro}
Data with correlated categorical responses arise frequently in applications. This may arise from units grouped into clusters (clustered data) or multiple measurements taken on the same unit (longitudinal data). For instance, we might expect the unemployment outcomes (employed, unemployed, not in labour force) of residents living in the same region to be correlated, due to similar job opportunities and socioeconomic levels. Ignoring the correlation structure and assuming that all observations are independent by fitting an ordinary multinomial regression model may result in biased estimates and inaccurate predictions. Multinomial mixed models can account for correlation by using group level random effects \cite[]{Daniels1997,Hartzel2001multinomial,Hedeker2003amixed-effects}. \\ 

For multinomial mixed models, it is a common practice to assume a multivariate normal distribution for the random effects. The multivariate normal distribution is easy to interpret and is convenient when we want to build more complicated correlation structures into our model. However, the resulting likelihood involves multidimensional integrals that cannot be solved analytically. The computational effort to evaluate the likelihood increases with the number of groups and categories, making it not suitable for large scale applications. In fact, \cite{Lee2016conjugate} showed that closed-form likelihoods for multinomial mixed models do not exist regardless of the random effect distribution, except for the special case when there are no covariates. \\

Various methods have been proposed to circumvent the computational obstacle for fitting multinomial mixed models. Among them are quadrature \cite[]{Hartzel2001multinomial,Hedeker2003amixed-effects}, Monte Carlo EM algorithm, pseudo-likelihood approach \cite[]{Hartzel2001multinomial} and Markov Chain Monte Carlo methods \cite[]{Daniels1997}. \cite{Jain1994arandom} proposed a random effects estimation approach using a discrete probability distribution approximation. Simulation based methods such as method of simulated moments \cite[]{McFadden1989amethod} and method of simulated maximum likelihood \cite[]{Gong2004economic,Hann2006estimation} are widely used in the  econometrics literature. Recently, \cite{Perry2016fast} proposed a fast moment-based estimation method that scales well for large samples and  which arguably can be extended for fitting multinomial mixed models. \cite{Kuss2007} used the fact that the multinomial model is a multivariate binary model and exploited a procedure proposed by \cite{Wright1998} for model fitting. Their approach has been criticized by \cite{Rooij2012} as they failed to realize that a multivariate link function is needed in the context of multinomial models. An alternative strategy using clustered bootstrap was subsequently proposed by \cite{Rooij2012}. Although some authors have considered the Dirichlet-multinomial model that results in a closed-form likelihood, it does not allow the incorporation of individual level covariates. \\


\cite{Chen2001anote} advocate using Poisson log-linear or non-linear mixed models, both with random effects, as surrogates to multinomial mixed models. Their method capitalizes on existing mixed models software packages for fitting generalized linear models with random effects. This allows multinomial mixed models to be fitted using an approximate likelihood from the Poisson surrogate models. Their results are based on extensions of the well known ``Poisson Trick'' \cite[]{Baker1994,Mccullagh1989generalized,Venables2002modern} that relates multinomial models with Poisson models via a respecification of the model formulae. Although clever, their methods have been criticized for being computationally inefficient \cite[]{Malchow-Moller2003estimation,Kuss2007} and sometimes producing nonsense results \cite[]{Kuss2007}. This might be due to the intractable likelihoods of their models and the various approximation methods being used in different software packages. The considerable execution time is especially problematic, where it can take up to months to fit the model to a moderate-sized  dataset \cite[]{Malchow-Moller2003estimation}! In this article, we propose a new approach based on an extension of the Gamma-Poisson model \cite[]{Lee2017sufficiency}, where the likelihood can be expressed in closed-form. Using the proposed estimation procedure, the parameters can be estimated via readily available packages for fitting generalized linear models. \\

The remaining paper is organized as follows. Section \ref{sec:ordinaryPoissonTrick} reviews the ``Poisson Trick'' for multinomial regression models with independent responses, and suggests a parsimonious fitting strategy when all covariates are categorical. In Section \ref{sec:extendedPoissonTrick} we propose a new approach for approximating the likelihood of multinomial regression models with random effects. The empirical performance of the proposed model is demonstrated via a simulation study and a reanalysis of the yogurt brand choice dataset as discussed by \cite{Chen2001anote} in Section \ref{sec:application}. Finally, we conclude with a discussion and a summary of our findings in Section \ref{sec:conclusion}.

\section{"Poisson Trick" for Independent Multinomial Responses} \label{sec:ordinaryPoissonTrick}

This section describes the relationship between multinomial and Poisson regression models for independent responses, which we refer to as the ``Poisson Trick''. The results are based on the well known fact that given the sum, Poisson counts are jointly multinomially distributed \cite[]{Mccullagh1989generalized}. 

\subsection{Derivation} \label{subsec:derivation_poissontrick}
Let $Y_{j} = (Y_{jq})_{q=1}^{Q}$ be the $Q \times 1$ response vector for observation $j$ with the corresponding probability vector $p_{j} = (p_{jq})_{q=1}^{Q}$, where $q$ indexes the multinomial category. A common approach to satisfy the two characteristics of probability (i) $0 \leq p_{jq} \leq 1$ for all $j$ and $q$; and (ii) $\sum_{q=1}^Q p_{jq} = 1$ is via\begin{equation} \label{eqn:pr}
p_{jq} = \zeta_{jq}/\zeta_{j+},
\end{equation} 
where $\zeta_{jq}$ is a positive user-specified function of covariates $x$ and fixed effects $\gamma$, and $\zeta_{j+} = \sum_{q=1}^Q \zeta_{jq}$. Depending on the type of variable under consideration, $x$ and $\gamma$ can be indexed by various combinations of $j$ and $q$ \cite[pp. 7-8]{mlogit_vignette}. Conditional on the multinomial sums $Y_{j+} = \sum_{q=1}^Q Y_{jq}$, $Y_js$ are independently multinomially distributed for each $j$, i.e.
\begin{equation}
Y_j|Y_{j+} \sim \mathcal{M}\left(Y_{j+},p_j\right) .
\end{equation} 

In multinomial models, $Y_{j+} = y_{j+}$ is treated as fixed. Suppose we instead treat $Y_{j+}$ as random and assume 
\begin{equation}
Y_{j+} \sim \mathcal{P}(\delta_j \zeta_{j+}),
\end{equation}
independently for each $j$. This results in a multinomial-Poisson mixture with the following joint probability function for each $j$:
\begin{align}
\text{P}(Y_j = y_j \cap Y_{j+}=y_{j+}) &=  \text{P}(Y_{j+}=y_{j+}) \text{P}(Y_j = y_j| Y_{j+}=y_{j+}) \nonumber \\ 
&= e^{-\delta_j\zeta_{j+}} \frac{(\delta_j\zeta_{j+})^{y_{j+}}}{y_{j+}!} \times \frac{y_{j+}!}{\prod_q y_{jq}!} \prod_q \left(\frac{\zeta_{jq}}{\zeta_{j+}}\right)^{y_{jq}} \nonumber \\
&= \prod_q \left\lbrace \frac{e^{-\delta_j \zeta_{jq} } \left(\delta_j  \zeta_{jq}\right)^{y_{jq}}}{y_{jq}!} \right\rbrace    \ \ \text{iff} \ \  Y_{j+} = y_{j+}.
\end{align} 

The marginal probability of $Y_j$ can then be obtained by summing the joint probability over all possible values of $Y_{j+}$:\\
\begin{align} \label{eqn:Pois_marginallik}
\text{P}(Y_j=y_j) &= \sum_{Y_{j+}=0}^\infty \prod_q \left\lbrace \frac{e^{-\delta_j \zeta_{jq}} (\delta_j \zeta_{jq})^{y_{jq}}}{y_{jq}!} \right\rbrace \nonumber \\
&= \prod_q \left\lbrace \frac{e^{-\delta_j \zeta_{jq}} (\delta_j  \zeta_{jq})^{y_{jq}}}{y_{jq}!} \right\rbrace. 
\end{align}  \

Thus, allowing the multinomial sums to be random according to a Poisson distribution results in
\begin{equation}
Y_{jq} \sim \mathcal{P}(\delta_j \zeta_{jq}),
\end{equation}
\noindent independently for each $j$ and $q$. Summing over all observations, the log-likelihood is
\begin{equation} \label{eqn:loglik_comparison_withoutRE}
\sum_j \ell^\mathcal{P}(\delta_j\zeta_{j+};Y_{j+}) + \sum_j \ell^\mathcal{M}\left(\zeta_j;Y_j|Y_{j+}\right) = \sum_j \sum_q \ell^\mathcal{P}(\delta_j\zeta_{jq}; Y_{jq}),
\end{equation}
where $\ell^\mathcal{P}$ and $\ell^\mathcal{M}$ denote the Poisson and multinomial log-likelihood functions respectively, and $\zeta_j = (\zeta_{jq})_{q=1}^Q$. The second term on the left hand side is the model we would like to fit, and the term on the right hand side is the model we actually fit. \\

To show that the Poisson surrogate model is an exact fit to the multinomial model, first note the log-likelihood corresponding to the multinomial model is
\begin{equation} \label{eqn:Multlik}
\sum_j \log(y_{j+!}) - \sum_j\sum_q \log(y_{jq}!) + \sum_j\sum_q y_{jq} \log\zeta_{jq} - \sum_j y_{j+} \log\zeta_{j+},
\end{equation}
and the log-likelihood of the Poisson surrogate model is
\begin{equation} \label{eqn:Poislik}
- \sum_j\delta_j \zeta_{j+} + \sum_j y_{j+} \log\delta_j + \sum_j\sum_q y_{jq} \log\zeta_{jq} - \sum_j \log(y_{j+}!).
\end{equation} \

Differentiating Equation \ref{eqn:Poislik} with respect to $\delta_j$ and setting it to $0$, we obtain $\hat{\delta}_j = y_{j+}/\zeta_{j+}$. Plugging in the maximizing value of $\delta_j$ into Equation \ref{eqn:Poislik}, we have
\begin{equation} \label{eqn:Poislik_deltahat}
-\sum_j y_{j+} + \sum_j y_{j+} \log y_{j+} - \sum_j y_{j+} \log\zeta_{j+} + \sum_j\sum_q y_{jq} \log \zeta_{jq}.
\end{equation} 

Equation \ref{eqn:Poislik_deltahat} is identical to Equation \ref{eqn:Multlik}, up to an additive constant. It follows that the maximum likelihood estimates, their asymptotic variances and tests for the fixed effects can be \textit{exactly} recovered under the Poisson surrogate model \cite[]{Richards1961}. That is, likelihood inference for $\zeta_{jq}$ is the same whether we regard $Y_{j+}$ as fixed (multinomial) or randomly sampled from independent Poissons. This result applies to the fixed effects model, with any parameterization of $\zeta_{jq}$, including:
\begin{itemize}
	\item Exponential transformations of linear combinations of \textit{categorical variables} and regression coefficients \cite[]{Mccullagh1989generalized,Agresti2013categorical},
	\item Exponential transformations of linear combinations of \textit{continuous variables} and regression coefficients,
	\item \textit{Any} monotonic transformations of linear combinations of covariates and regression coefficients,
	\item Nonlinear functions of covariates and regression coefficients,
	\item Nonparametric formulations.
\end{itemize} \

The Poisson surrogate model eliminates $\zeta_{j+}$ from the denominator of the multinomial probabilities. This makes sense intuitively, as we do not expect the multinomial sums to provide any useful information in estimating the fixed effects.  
Given that $\hat{\delta}_{j}$ can also be obtained by setting the fitted values of the multinomial sums $\hat{Y}_{j+}=\text{E}(Y_{j+})$ equal to the observed counts $y_{j+}$ in the Poisson surrogate model, $\delta_j$ has the effect of recovering the multinomial sums. The key idea is to include a separate constant $\delta_j$ for each unique combination of covariates in the Poisson surrogate models.

\subsection{Specifying Model Formulae for Poisson Surrogate Models}
\label{subsec:modelformula_PoisTrickOrdinary}
For purposes of exposition, the model formulae in this section are written in terms of the \textsf{R} language \cite[]{R2017}, although this article is not concerned with software packages per se. Multinomial models are fitted using the multinom() function within the nnet package \cite[]{nnet}; Poisson models are fitted using the glm() function within the stats package.\\

For concreteness, consider the \textit{non-parallel baseline category logit models}. The ``baseline category logit'' assumption refers to the following: treating category $1$ as the baseline category with $\zeta_{j1} = 1 \ \forall j$ without loss of generality, we model $\log(p_{jq}/p_{j1})$ = $\log(\zeta_{jq})$ as a linear function of covariates $x$ and regression coefficients $\gamma$. This assumption is not necessary, but chosen so that the model formulae can be illustrated using functions within the stats package. The ``non-parallel'' assumption refers to covariate effects that vary across categories \cite[]{Fullerton2016ordered}, i.e. all elements of the $\gamma$ vector are indexed by $q$. That is, if the set of logits are plotted against the covariate on the same graph, a set of straight lines with slopes that are in general different will be obtained. Later we shall discuss cases where we relax this assumption. \\

Consider a hypothetical dataset with two predictors $X_1$ and $X_2$ (these can be categorical or continuous) and a multinomial outcome vector $Y$ with $Q=3$ categories. In \textit{short format}, each row of data represents an observation with a $3$-dimensional outcome vector $(Y_1,Y_2,Y_3)$. Poisson models treat the outcomes of each observation as independent and glm() requires data to be presented in \textit{long format}. This requires an additional factor $C$ that denotes the category memberships. Each row now comprises a scalar outcome, resulting in $3$ rows of data per observation. The first few rows of data in both short and long format are shown in Table \ref{tab:multinomialdata}. \\

\begin{table}[H]  
	\caption{Multinomial data.}  \label{tab:multinomialdata}
	
	\begin{subtable}{.5\linewidth} 
		\centering
		\caption{Short format.}
		\begin{tabular}{cccccc}
			\toprule \midrule
			\multicolumn{1}{c}{Obs} & \multicolumn{1}{c}{$X_1$} & \multicolumn{1}{c}{$X_2$} & \multicolumn{1}{c}{$Y_1$} & \multicolumn{1}{c}{$Y_2$} & \multicolumn{1}{c}{$Y_3$} \\ \midrule
			1 & 0 & 0 & 3 & 5 & 2 \\ 
			2 & 0 & 1 & 5 & 5 & 0 \\
			3 & 1 & 0 & 7 & 2 & 1\\
			4 & 1 & 1 & 1 & 3 & 6 \\
			-- & -- & -- & -- & -- & -- \\ 	\bottomrule
		\end{tabular}
	\end{subtable}%
	\begin{subtable}{.5\linewidth}
		\centering
		\caption{Long format.}
		\begin{tabular}{ccccc}
			\toprule \midrule
			\multicolumn{1}{c}{I} & \multicolumn{1}{c}{$X_1$} & \multicolumn{1}{c}{$X_2$} & \multicolumn{1}{c}{C} & \multicolumn{1}{c}{Y} \\ \midrule
			1 & 0  & 0  & 1  & 3 \\
			1 & 0  & 0  & 2  & 5 \\
			1 & 0  & 0  & 3  & 2 \\
			2 & 0  & 1  & 1  & 5 \\ 
			2 & 0  & 1  & 2  & 5 \\
			2 & 0  & 1  & 3  & 0 \\
			-- & --  & --  & --  & -- \\  \bottomrule
		\end{tabular}
	\end{subtable}  
\end{table}

Table \ref{tab:ordinaryPoisTrick_generic} shows the equivalant relationship between non-parallel multinomial models and the corresponding Poisson models, where the parameters satisfy the usual constraints for identifiability. The Poisson surrogate models possess several important features:
\begin{enumerate}
	\item The model includes an indicator variable $I$ (that corresponds to $\log\delta_j$ in Section \ref{subsec:derivation_poissontrick}) for each observation, although this can be simplified when all covariates are categorical. 
	This is to ensure the exact recovery of the multinomial sums, as the fixed sums are treated as random in the Poisson models. As a result, we do not interpret the coefficients of $I$ since they are just nuisance parameters. 
	\item The category membership indicator $C$ enters as a covariate in the Poisson models, where the coefficients correspond to the intercepts in the multinomial modelsv so that the counts are allowed to vary by category.
	\item The model includes interaction terms between $X$ and $C$ (denoted by $*$ in the model formula), where the coefficients correspond to the slopes in the multinomial models. This is due to the non-parallel assumption where each category has a separate slope, and also the fact that multinomial models treat the response counts jointly for each observation, whereas Poisson models treat each response count as a separate observation. It is important that these interaction terms are included even if they are not significant. For multinomial models where some (\textit{partial models}) or all (\textit{parallel models}) of the covariate effects do not vary across categories, the equivalent Poisson models can be obtained by modifying the interaction structure between $X$ and $C$ accordingly. For instance, in parallel models where all categories share the same covariate effects, there is no need to include the interation terms between $X$ and $C$, since the slopes do not vary across categories. 
\end{enumerate} 

\begin{table}[H]
	\centering
	\caption{Equivalent relationship between non-parallel multinomial models and the corresponding Poisson models.}
	\label{tab:ordinaryPoisTrick_generic}
	\begin{threeparttable}
	\begin{tabular}{ll}
		\toprule\midrule
		\textbf{Multinomial$^1$}   & \textbf{Poisson$^2$}     \\    \midrule
		$Y \sim 1$  &  $Y \sim I + C$           \\
		$Y \sim X_1$  &   $Y \sim I + C + C*X_1$           \\
		$Y \sim X_1 + X_2$  &  $Y \sim I + C + C*X_1 + C*X_2$         \\
		$Y \sim X_1 + X_2 + X_1*X_2$  &   $Y \sim  I + C + C*X_1 + C*X_2 + C*X_1*X_2$  \\   \bottomrule\addlinespace[1ex]
	\end{tabular} 
	\begin{tablenotes}\footnotesize
		\item[$^1$] Syntax for using multinom() in \textsf{R}, where data are presented in short format and $Y$ is a vector of response counts.
		\item[$^2$] Syntax for using glm() in \textsf{R}, where data are presented in long format and $Y$ is a scalar response count.
	\end{tablenotes}
	\end{threeparttable}
\end{table} 

When writing the model formula, it is important to specify $I$ and $C$ as \textit{factors} due to their categorical nature. This can be achieved via the factor() function in \textsf{R}. \\

\noindent \textbf{Special Case: Categorical Covariates} \\
When all covariates are categorical, the model formulae in Table \ref{tab:ordinaryPoisTrick_categorical} offer a more parsimonious option for fitting the Poisson models without having to estimate a separate parameter for each observation. 

\begin{table}[H]
	\centering
	\caption{Equivalent relationship between non-parallel multinomial models and the corresponding Poisson models, when all the covariates are categorical.}
	\label{tab:ordinaryPoisTrick_categorical}
	\begin{threeparttable}
	\begin{tabular}{ll}
		\toprule\midrule
		\setcounter{footnote}{0}
		\textbf{Multinomial$^1$}   & \textbf{Poisson$^2$}     \\    \midrule
		$Y \sim 1$     &  $Y \sim C$           \\
		$Y \sim X_1$  &   $Y \sim X_1 + X_1*C$           \\
		$Y \sim X_1 + X_2$  &  $Y \sim X_1*X_2 + X_1*C  + X_2*C$         \\
		$Y \sim X_1 + X_2 + X_1*X_2$ &   $Y \sim X_1*X_2*C$  \\   \bottomrule\addlinespace[1ex] 
	\end{tabular} 
	\begin{tablenotes}\footnotesize
		\item[$^1$] Syntax for using multinom() in \textsf{R}, where data are presented in short format and $Y$ is a vector of response counts.
		\item[$^2$] Syntax for using glm() in \textsf{R}, where data are presented in long format and $Y$ is a scalar response count.
	\end{tablenotes}
	\end{threeparttable}
\end{table} 

As stated above, the key to achieving the 1-1 correspondence between multinomial and Poisson models (with the same link function) is to include a separate constant for each unique combination of covariates. For categorical covariates, this can be achieved by including the full interaction among the predictors in the Poisson model. When all the covariates are categorical, the interaction term has the precise effect of pooling groups of observations with identical covariates. Fitting such models is equivalent to fitting the observation index as a factor (Table \ref{tab:ordinaryPoisTrick_generic}), but the pooling results in a smaller effective data frame, and therefore smaller storage requirements and faster fitting speed, with no loss of information. Of course, if there are many factors, there may not be much saving, because it will be comparatively rare for different observations to have all the same factor level combinations. 



\section{Extending the ``Poisson Trick'' for Correlated Multinomial Responses}
\label{sec:extendedPoissonTrick}

\subsection{Derivation}
Consider a set of observations which fall into a collection of $I$ groups and let $\lambda_i = (\lambda_{iq})_{q=1}^{Q}$ be a vector-valued random effect for group $i$. Each observation belongs to only a single group. Extending the notation in Section \ref{sec:ordinaryPoissonTrick}, the $Q \times 1$ response vector for observation $j$ in group $i$ is $Y_{ij} = (Y_{ijq})_{q=1}^{Q}$, with the corresponding probability vector $p_{ij} = (p_{ijq})_{q=1}^{Q}$, where $p_{ijq} = \lambda_{iq}\zeta_{ijq}/\sum_{q=1}^Q \lambda_{iq}\zeta_{ijq}$. Conditional on the multinomial sums $Y_{ij+} = \sum_{q=1}^Q Y_{ijq}$ and the random effects $\lambda_{i}$, 
the counts are Multinomially distributed:
\begin{equation}
Y_{ij}|Y_{ij+},\lambda_i \sim \mathcal{M}\left(Y_{ij+}, p_{ij}\right).
\end{equation} \

In analogy to the results in Section \ref{sec:ordinaryPoissonTrick}, given the random effects, we treat $Y_{ij+}$ as random and assume
\begin{equation}
Y_{ij+}|\lambda_i \sim \mathcal{P}\left(\delta_{ij}\sum_{q=1}^Q \lambda_{iq}\zeta_{ijq}\right),
\end{equation}
\noindent independently for each $i$ and $j$. This gives
\begin{equation}
	Y_{ijq}|\lambda_{iq} \sim \mathcal{P}(\delta_{ij} \lambda_{iq} \zeta_{ijq}),
\end{equation}
\noindent independently for each $j$ and $q$. The probability argument in Equation \ref{eqn:loglik_comparison_withoutRE} still holds, now conditional on the random effects:
\begin{equation} \label{eqn:loglik_comparison_withRE}
\sum_i \sum_j \ell^\mathcal{P}\left(\delta_{ij}\sum_{q=1}^Q \lambda_{iq}\zeta_{ijq};Y_{ij+}|\lambda_i\right) + \sum_i \sum_j \ell^\mathcal{M}\left(\zeta_{ij};Y_{ij}|Y_{ij+},\lambda_i\right) = \sum_i \sum_j \sum_q \ell^\mathcal{P}\left(\delta_{ij}\lambda_{iq}\zeta_{ijq}; Y_{ijq}|\lambda_{iq}\right),
\end{equation} 
\noindent where  $\zeta_{ij} = (\zeta_{ijq})_{q=1}^Q$. If the random effects are observed, the conditional probability statement above imply a 1-1 exact correspondence between the multinomial and the Poisson surrogate models. However, due to the unobserved nature of the random effects, interest lies in the marginal distribution, obtained by integrating out the random effects. This results in an approximate relationship between the two models. It turns out that the marginal likelihood of the approximating Poisson surrogate model (right hand side of Equation \ref{eqn:loglik_comparison_withRE}) can be expressed in closed-form if we assume an independent Gamma model for the random effects, with $\text{E}(\lambda_{iq}) = \alpha_q/\beta_q$ and $\text{Var}(\lambda_{iq}) = \alpha_q/\beta_q^2$, i.e. $\lambda_{iq} \sim \mathcal{G}(\alpha_q,\beta_q)$ \cite[]{Lee2016conjugate}. \\

With this assumption for the distribution of the random effects, the marginal likelihood of the multinomial model that we would like to fit (second term on the left hand side of Equation \ref{eqn:loglik_comparison_withRE}) is given by
\begin{align} \label{eqn:lik_multinomMM}
L^{\mathcal{M}} &= \prod_i \int \cdots \int \prod_j 	 \left\lbrace 
\frac{y_{ij+}!}{\prod_q y_{ijq}!}
\prod_q \left(\frac{\lambda_{iq}\zeta_{ijq}}{\sum_{q=1}^Q \lambda_{iq}\zeta_{ijq}}\right)^{y_{ijq}}   \right\rbrace
\times
\prod_q\frac{\beta_q^{\alpha_q}\lambda_{iq}^{\alpha_q-1}e^{-\beta_q\lambda_{iq}}}
{\Gamma(\alpha_q)}
\ \text{d}\lambda_{i1}\ldots \text{d}\lambda_{iQ}.  
\end{align} 
\noindent This does not generally exhibit a closed-form solution regardless of the random effect distribution, unless in the special cases of no covariate or with only group specific covariates \cite[]{Lee2016conjugate}. Numerical or simulation methods can be used to approximate the likelihood, with computational efforts increasing with increasing number of groups and categories. On the other hand, the marginal likelihood of the Poisson surrogate model can be expressed in closed-form:
\begin{align} 
L^P &= \prod_i \left\lbrace \int \cdots \int \prod_j \prod_q	 
\frac{e^{-\delta_{ij}\lambda_{iq}\zeta_{ijq}}(\delta_{ij}\lambda_{iq}\zeta_{ijq})^{y_{ijq}}}
{y_{ijq}!}
\prod_q\frac{\beta_q^{\alpha_q}\lambda_{iq}^{\alpha_q-1}e^{-\beta_q\lambda_{iq}}}
{\Gamma(\alpha_q)}
\ \text{d}\lambda_{i1}\ldots \text{d}\lambda_{iQ}  \right\rbrace
\nonumber
\\
&= \prod_i \left\lbrace \prod_q\frac{\Gamma(\alpha_q+y_{i+q})\beta_q^{\alpha_q}}
{\Gamma(\alpha_q)(\beta_q+\sum_j\delta_{ij} \zeta_{ijq})^{\alpha_q+y_{i+q}}}
\times
\prod_j\prod_q\frac{(\delta_{ij} \zeta_{ijq})^{y_{ijq}}}{y_{ijq}!} \right\rbrace. \label{eqn:lik_poissonMM}
\end{align} 
\noindent The Poisson surrogate model is an extension of the Gamma-Poisson model as proposed by \cite{Lee2017sufficiency} and \cite{Lee2016conjugate} to allow the modelling of counts for multiple categories. \\

As a consequence of Equation \ref{eqn:lik_poissonMM}, we have 
\begin{equation} \label{eqn:expectedvalue_populationaveraged}
\text{E}(Y_{ijq}) = \frac{\alpha_q}{\beta_q} \delta_{ij} \zeta_{ijq}.
\end{equation} 
\noindent Refer to the appendix for details. This is the population-averaged expected value and is not suitable for prediction in general, as it does not take into account the cluster effect. However, it can be useful for out of sample prediction, when there are no samples present in a particular group. \\

\noindent \textbf{Special Case: Var($\boldsymbol{\lambda_{iq}}$) approaches $\boldsymbol{0}$} \\
When Var($\lambda_{iq}$) approaches $0$ for all $q$, the model reduces to the special case of no random effects as outlined in Section \ref{sec:ordinaryPoissonTrick}, and the exact correspondence between the multinomial and the Poisson models can be regained. 

\subsection{Identifiability} \label{subsec:identifiability}
There is some lack of identifiability with the model formulation given by Equation \ref{eqn:lik_poissonMM}, characterized by non-uniqueness of the maximum likelihood estimates. 
There is an identifiability issue between $\lambda_{iq}$ and $\delta_{ij}$, and also between $\lambda_{iq}$ and the category intercepts. To fix this, we impose the constraint of $\alpha_q = 1/\beta_q$ so that  $\text{E}(\lambda_{iq}) = 1$. As a consequence, $\lambda_{iq} \sim \mathcal{G}(1/\beta_q,\beta_q)$ and $\text{Var}(\lambda_{iq}) = \beta_q$. Also, we only require a random effect for each logit, and thus a constraint for the random effects associated with the baseline category $q=1$ is needed. Denote $u_{iq} = \log\lambda_{iq}$. Several authors such as \cite{Agresti2013categorical} (pp.514) and \cite{Hartzel2001multinomial} considered a multivariate normal distribution for the random effects, i.e. $(u_{iq})_{q=2}^Q \sim \mathcal{N}(0,\Sigma)$, where $\Sigma$ is a $Q-1$ by $Q-1$ variance-covariance matrix. This is equivalent to saying that $u_{i1}=0$ for all $i$, or $\sigma_{11} = 0$. The equivalent statement in our proposed model is to fix $\lambda_{i1} = 1$ for all $i$. This is tantamount to saying $\text{Var}(\lambda_{i1}) = \beta_1$ approaches $0$, and thus $\alpha_1$ approaches $\infty$. 

\subsection{Prediction of Random Effects and Fitted Values}
We focus on the \textit{best predictor} (BP) for random effects prediction, i.e. the predictor that minimises the overall mean squared error of prediction.  \cite{McCulloch2008generalized} shows that the BP is given by the posterior expectation of the random effect. Under the proposed Poisson surrogate model, the BP is given by 
\begin{equation}  \label{eqn:BP}
\text{BP}(\lambda_{iq}) = \hat{\lambda}_{iq} \equiv \underset{\lambda^\star}{\text{argmin}} \ \text{E}(\lambda_{iq}-\lambda^\star)^2 := \text{E}(\lambda_{iq}|y),
\end{equation}

\noindent which can be calculated via 
\begin{equation} \label{eqn:predictedRE_generalformula}
\hat{\lambda}_{iq} = \frac{\displaystyle \int_{-\infty}^{\infty} \lambda_{iq} f(\lambda_{iq}) f(y|\lambda_{iq}) \ d\lambda_{iq}}{\displaystyle \int_{-\infty}^{\infty} f(\lambda_{iq}) f(y|\lambda_{iq}) \ d\lambda_{iq}} \ .
\end{equation} \

\noindent Solving for the integral, the BP is
\begin{equation}
\hat{\lambda}_{iq}  = \dfrac{Y_{i+q} + 1/\beta_q}{\sum_j  \delta_{ij} \zeta_{ijq} + \beta_q},
\end{equation}
\noindent where $Y_{i+q} = \sum_j Y_{ijq}$. $\hat{\lambda}_{iq}$ depends on the parameters $\delta_{ij}$, $\gamma$ and $\beta_q$, in which we replace by their estimators, leading to the \textit{empirical best predictor} (EBP). The fitted values can then be defined as 
\begin{equation} \label{eqn:fittedPoisCGLMM}
\hat{Y}_{ijq} = \delta_{ij} \hat{\lambda}_{iq} \zeta_{ijq},
\end{equation}
where we replace $\delta_{ij}$ and $\zeta_{ijq}$ by their respective estimators $\hat{\delta}_{ij}$ and $\hat{\zeta}_{ijq}$.

\subsection{Parameter Estimation} \label{subsec:ECM}
Consider the parameterization $\zeta_{ijq} = \exp(\eta_{ijq})$ which is widely adopted in practice, where $\eta_{ijq} = x_{ijq}^T\gamma$, where $x_{ijq}$ and $\gamma$ are both vectors. The chosen index structure for $x$ and $\gamma$ encompasses a variety of possible scenarios: (i) category-specific predictors with generic coefficients $x_{ijq}^T\gamma$, (ii) category-specific predictors with category-specific coefficients $x_{ijq}^T\gamma_q$, and (iii) observation-specific predictors with category-specific coefficients $x_{ij}\gamma_q$. This can be achieved by creating the appropriate interaction terms between the predictor and the category indicator variable, thus modifying the model matrix. Note that observation-specific predictors must be paired with choice-specific coefficients. Otherwise they will disappear in the differentiation when we consider the log-odds. \\

Denote $\theta = (\gamma,(\beta_q)_{q=2}^Q)$, where $\gamma$ includes the incidental parameters $\log(\delta_{ij})$ for all $i$ and $j$. Algorithm \ref{alg:ECM_PoisSurrogate} presents an Expectation/Conditional Maximization (ECM) algorithm \cite[]{Meng1993Maximum} for parameter estimation of the Poisson surrogate model. Refer to the appendix for a detailed derivation.

\begin{algorithm}[h!]
	\begin{center}
		\begin{minipage}[t]{150mm}
			\hrule
			\vskip4mm\noindent
			\textbf{Initialize} $\theta$. \\ 
			\textbf{Cycle}: \\
			\While{relative differences in the parameter estimates are not negligible}{
				\textbf{E-Step}: Calculate for each $i$ and $q$:
				\begin{gather*}
				\hat{\lambda}_{iq}^{(t+1)} = \text{E}\left(\lambda_{iq}\middle|(y_{ijq})_j,\theta^{(t)}\right) = \frac{y_{i+q}+1/\beta_q^{(t)}}{\sum_j e^{x_{ijq}\gamma^{(t)}}+1/\beta_q^{(t)}} \\
				\hat{\chi}_{iq}^{(t+1)} = \text{E}\left(\log(\lambda_{iq})\middle|(y_{ijq})_j,\theta^{(t)}\right) = \psi\left(y_{i+q}+1/\beta_q^{(t)}\right) - \log\left(\sum_j e^{x_{ijq}\gamma^{(t)}}+1/\beta_q^{(t)}\right),
				\end{gather*} 
				where $\psi(\cdot)$ is the digamma function.\\
				\textbf{CM-Step}: 
				\begin{itemize}
					\item Obtain $\gamma^{(t+1)}$ by fitting a Poisson log-linear model with $y_{ijq}$ as the response and $X_{ijq}$ as the design matrix, with $\hat{\lambda}_{iq}^{(t+1)}$ as offset. $X_{ijq}$ includes indicator variables for each unique combination of covariates.
					\item Obtain $\beta_q^{(t+1)}$ for each $\beta_q$ for $q=2$ to $Q$ by maximizing
					\begin{equation*}
					\sum_i \left\lbrace (1/\beta_q-1)\hat{\chi}_{iq}^{(t+1)} - \hat{\lambda}_{iq}^{(t+1)}/\beta_q - \log(\beta_q)/\beta_q - \log\Gamma(1/\beta_q)
					\right\rbrace,
					\end{equation*}
					where $\Gamma(\cdot)$ is the gamma function.
				\end{itemize}
			}
			\hrule
			\vskip3mm
		\end{minipage}
	\end{center}
	\caption{\it Expectation/Conditional Maximization (ECM) algorithm for fitting the Poisson surrogate model.} 
	\label{alg:ECM_PoisSurrogate}
\end{algorithm}


\section{Yogurt Brand Choice Dataset} \label{sec:application}
We consider the yogurt brand choice dataset previously analyzed by \cite{Jain1994arandom} and \cite{Chen2001anote}. \cite{Jain1994arandom} approximated the likelihood of a multinomial logit model with Gaussian random effects using a discrete distribution. \cite{Chen2001anote} approximated the multinomial logit model using the Poisson log-linear model and Poisson nonlinear model, both with Gaussian random effects. \\

The dataset consists of purchases of yogurt by a panel of $100$ households in Springfield, Missouri, and were originally provided by A. C. Nielsen. The data
were collected by optical scanners for about two years and correspond to $2,412$ purchases. Variables collected include brand, price and presence of newspaper feature advertisements for each purchase made by households in the panel. Price and feature advertisements are choice-specific variables. We assume a \textit{parallel baseline logit} model by assigning generic coefficients $\gamma$ to these variables, as we do not expect the effect of price and feature advertisements on the probability of purchase to vary according to brands. The four brands of yogurt: Yoplait, Dannon, Weight Watchers, and Hiland account for market shares of 34\%, 40\%, 23\%, and 3\% respectively. Following \cite{Chen2001anote}, we put Hiland as the reference brand. Table \ref{tab:yogurtdata} presents the yogurt data in both long and short format. The letters `f' and `p' represent the feature and price variables respectively, with the letter that follows denoting the brand. For instance, `fy' stands for `feature of Yoplait' and `pd' stands for `price of Dannon'. \\

\begin{table}
	\caption{Yogurt data.}  \label{tab:yogurtdata}
	
	\begin{subtable}{1\textwidth} 
		\centering
		\caption{Short format.}
			\begin{tabular}{cccccccccccccc} 
				\toprule\midrule
				\multicolumn{1}{c}{id} & \multicolumn{1}{c}{obs} & \multicolumn{1}{c}{yoplait} & \multicolumn{1}{c}{dannon} & \multicolumn{1}{c}{weight} & \multicolumn{1}{c}{hiland} & \multicolumn{1}{c}{fy} &
				\multicolumn{1}{c}{fd} &  \multicolumn{1}{c}{fw} &
				\multicolumn{1}{c}{fh} &  \multicolumn{1}{c}{py} &
				\multicolumn{1}{c}{pd} & \multicolumn{1}{c}{pw} &
				\multicolumn{1}{c}{ph}\\ \midrule
				1 & 1 & 0 & 1 & 0 & 0 & 0 & 0 & 0 & 0 & 0.108 & 0.081 &0.079 & 0.061 \\ 
				1 & 2 & 0 & 1 & 0 & 0 & 0 & 0 & 0 & 0 & 0.108 & 0.098 &0.075 & 0.064 \\
				1 & 3 & 0 & 1 & 0 & 0 & 0 & 0 & 0 & 0 & 0.108 & 0.098 &0.086 & 0.061 \\ 
				-- & -- & -- & -- & -- & -- & -- & -- & -- & -- & -- & -- &-- & -- \\ 
				2 & 9 & 1 & 0 & 0 & 0 & 0 & 0 & 0 & 0 & 0.108 & 0.098 &0.079 & 0.050 \\ 
				-- & -- & -- & -- & -- & -- & -- & -- & -- & -- & -- & -- &-- & -- \\ 
				100 & 2412 & 0 & 0 & 1 & 0 & 0 & 0 & 0 & 0 & 0.108 & 0.086 &0.079 & 0.043 \\ 
				\bottomrule
			\end{tabular}
	\end{subtable}%
	
	\bigskip
	
	\begin{subtable}{1\textwidth}
		\centering
		\caption{Long format.}
			\begin{tabular}{cccccc}
				\toprule\midrule
				\multicolumn{1}{c}{id} & \multicolumn{1}{c}{obs} & \multicolumn{1}{c}{feature} & \multicolumn{1}{c}{price} & \multicolumn{1}{c}{count} & \multicolumn{1}{c}{brand} \\ \midrule
				1 & 1 & 0  & 0.108  & 0  & yoplait \\
				1 & 1 & 0  & 0.081  & 0  & dannon \\
				1 & 1 & 0  & 0.079  & 1  & weight \\
				1 & 1 & 0  & 0.061  & 0  & hiland \\ 
				1 & 2 & 0  & 0.108  & 0  & yoplait \\
				1 & 2 & 0  & 0.098  & 1  & dannon \\
				-- & -- & --  & --  & --  & -- \\
				100 & 2412 & 0  & 0.043  & 1  & hiland \\
				\bottomrule
			\end{tabular} 
	\end{subtable}  
\end{table}

We fit the models proposed in Sections \ref{sec:ordinaryPoissonTrick} and \ref{sec:extendedPoissonTrick}, and compare our results to that of \cite{Chen2001anote}, fitted using the SAS macro GLIMMIX and the SAS procedure NLMIXED. The results are presented in Table \ref{tab:results}. The preference ordering of the brands are the same for all models, i.e. Yoplait is the most preferred brand, followed by Dannon, Weight Watchers and Hiland. The slope parameters estimates have the expected signs for all models. An increase in price is associated with a decrease in the probability of purchase. Feature advertisement tends to increase the chance of purchase. The household-to-household variation in the probability of purchase for Weight Watchers is much larger than the other brands, although none are significant (p > 0.05). \\

In comparing the estimates between models, we note that the fixed effects model is likely to produce biased estimates as it did not take into account of the correlation induced by multiple purchases from the same household. The parameter estimates of NLMIXED and Gamma-Poisson are uniformly larger than that of GLIMMIX, except for the intercept associated with Weight Watchers (for NLMIXED) and for the slope associated with price (for Gamma-Poisson). The estimates of the standard errors from NLMIXED and Gamma-Poisson are also uniformly larger than that of GLIMMIX. These differences can be attributed to the different distributional assumptions of the random effects, and also the different approximations used in GLIMMIX and NLMIXED to estimate the intractable likelihood. In this regard, our model exhibit a closed-form likelihood that allows exact inference to be performed with respect to the approximating model. \\

We tried to fit a simplified version of GLIMMIX using the glmer() function within the lme4 package in \textsf{R}, with just a random effect per household (ignoring the choice effect). However, the model failed to converge within a few months, even though \cite{Chen2001anote} claimed that the GLIMMIX model coverged in SAS.

\begin{table}[h]
	\centering
	\caption{Regression estimates for the yogurt data, and the associated standard errors.}
	\label{tab:results}
	\begin{threeparttable}
	\resizebox{\textwidth}{!}{%
	\begin{tabular}{lcccc}          
		\toprule\midrule
           &   &  \multicolumn{3}{c}{\textbf{Random Effects}} \\
		\cline{3-5}
		\textbf{Variables}  & \textbf{Fixed Effects$^1$} & \textbf{GLIMMIX$^2$} & \textbf{NLMIXED$^3$}& \textbf{Gamma-Poisson$^4$} \\
		\midrule
		Dannon     & 3.716 (0.145)   & 3.838 (0.231)    & 4.130 (0.648)    & 4.616 (0.309)   \\
		Weight     & 3.074 (0.145)   & 2.242 (0.241)    & 1.046 (0.671)    & 3.677 (0.392)   \\
		Yoplait    & 4.450  (0.187)  & 4.626 (0.261)    & 4.805 (0.699)    & 5.275 (0.342)   \\
		Feature    & 0.491 (0.120)   & 0.730 (0.121)    & 0.956 (0.185)    & 0.785 (0.178)   \\
		Price      & -36.658 (2.437) & -40.012 (2.562)  & -36.686 (3.725)  & -40.881 (3.778) \\
		$\beta_{Dannon}$  & N/A             & N/A              & N/A              & 2.203 (0.134)   \\
		$\beta_{Weight}$  & N/A             & N/A              & N/A              & 6.067 (0.374)   \\
		$\beta_{Yoplait}$  & N/A             & N/A              & N/A              & 1.918 (0.135)   \\
		\bottomrule\addlinespace[1ex]
	\end{tabular}}
	\begin{tablenotes}\footnotesize
		\item[$^1$] Fitted using the glm() function in \textsf{R}, using the ``Poisson Trick'' as outlined in Section \ref{sec:ordinaryPoissonTrick}.
		\item[$^2$] Poisson log-linear model with Gaussian random effects, fitted by \cite{Chen2001anote} using the SAS macro GLIMMIX.
		\item[$^3$] Poisson nonlinear model with Gaussian random effects, fitted by \cite{Chen2001anote} using the SAS procedure NLMIXED.
		\item[$^4$] Poisson log-linear model with Gamma (multiplicative) random effects fitted using the ECM algorithm, as outlined in Section \ref{sec:extendedPoissonTrick}. 
	\end{tablenotes}
	\end{threeparttable}
\end{table}


\section{Concluding Remarks} \label{sec:conclusion}
In this article, we presented methods for fitting various multinomial regression models via the so-called ``Poisson Trick'' and its extensions. The ``Poisson Trick'' for fitting fixed effects multinomial regression models is handy when the direct fitting of multinomial models is not supported, for instance the INLA package \cite[]{Rue2009approximate} in \textsf{R}. For multinomial regression models with random effects, there exist a variety of experience for using the existing extensions proposed by \cite{Chen2001anote}, from taking months to fit a moderate sized dataset \cite[]{Malchow-Moller2003estimation}, producing nonsense results \cite[]{Kuss2007} to non-convergence in our experience of fitting the yogurt brand choice dataset. We proposed an extension of the ``Poisson Trick'' using Gamma (multiplicative) random effects. In contrast to the models by \cite{Chen2001anote}, our model exhibits a closed-form likelihood and can be maximized using existing functions for fitting generalized linear models that are stable and heavily optimized, without having to approximate the integrals.




\section{Appendix}



\subsection{Derivation of the Population-Averaged Expected Values in Equation \ref{eqn:expectedvalue_populationaveraged}}
Equation \ref{eqn:lik_poissonMM} is also equivalent to
\begin{equation} \label{eqn:Poisson2}
\prod_i \left\lbrace \prod_q  \left[ \frac{\Gamma(\alpha_q+y_{i+q})}
{\Gamma(\alpha_q) y_{i+q}!} \left(  \frac{\sum_j \delta_{ij} \zeta_{ijq}}{\beta_q+\sum_j\delta_{ij} \zeta_{ijq}} \right) ^{y_{i+q}}   
\left( \frac{\beta_q}{\beta_q+\sum_j\delta_{ij}\zeta_{ijq}}  \right)^{\alpha_q}  \right]  \times  \prod_q  \left[ \frac{y_{i+q}!}{\prod_j y_{ijq}!}  \frac{\prod_j (\delta_{ij} \zeta_{ijq})^{y_{ijq}}}{(\sum_{j} \delta_{ij} \zeta_{ijq})^{y_{i+q}}} \right] \right\rbrace.
\end{equation} \

This results in two different interpretations for the extended Gamma-Poisson surrogate model: 
\begin{enumerate}
	\item For each $i$ and $q$, $Y_{ijq}$ is independent negative multinomial $\mathcal{NM}\left(\alpha_q, \frac{\delta_{ij} \zeta_{ijq}}{\beta_q+ \sum_j \delta_{ij} \zeta_{ijq}}   \right)$ (Equation \ref{eqn:lik_poissonMM}). 
	\item For each $i$ and $q$, the category sums  $Y_{i+q}$ are independent Negative Binomial \\ $\mathcal{NB}\left(\alpha_q,\frac{\sum_j\delta_{ij} \zeta_{ijq}}{\beta_q+\sum_j\delta_{ij} \zeta_{ijq}}\right)$, and conditional on the $Y_{i+q}$, $Y_{ijq}$ is independent multinomial $\mathcal{M}\left(Y_{i+q},  \frac{\delta_{ij} \zeta_{ijq}}{\sum_{j} \delta_{ij} \zeta_{ijq}}  \right)$ (Equation \ref{eqn:Poisson2}). 
\end{enumerate} \

Taking expectation of both Equations \ref{eqn:lik_poissonMM} and \ref{eqn:Poisson2} with respect to $Y_{ijq}$ gives rise to the population-averaged expected value given in Equation \ref{eqn:expectedvalue_populationaveraged}. The definitions of negative multinomial and negative binomial distributions are given in the following subsections.

\subsubsection{Negative Multinomial Distribution}
This is the distribution on the $n+1>2$ non-negative integers outcomes $\lbrace X_0,\dots,X_n \rbrace$, with corresponding probability of occurence $p = \lbrace p_0,\dots,p_n \rbrace$ and probability mass function
\begin{equation*}
\Gamma\left( \sum_{i=0}^n x_i \right) \frac{p_0^{x_0}}{\Gamma(x_0)} \prod_{i=1}^n \frac{p_i^{x_i}}{x_i!},
\end{equation*}

\noindent for parameters $x_0 > 0$ and $p=(p_i)_{i=1}^n$, where $p_i \in (0,1)$ for all $i$, $\sum_{i=0}^n p_i = 1$ and $\Gamma(\cdot)$ is the Gamma function. We write $Y \sim \mathcal{NM}(x_0,p)$. For positive integer $x_0$, the negative multinomial distribution can be recognized as the joint distribution of the n-tuple $\lbrace X_1,\dots,X_n \rbrace$ when performing sampling until $X_0$ reaches the predetermined value $x_0$. The mean vector of negative multinomial distribution is given by $\frac{x_0}{p_0} p$.

\subsubsection{Negative Binomial Distribution}
This is the distribution on the non-negative integers outcome $X$, with corresponding probability of occurence $p$ and probability mass function
$$
\frac{\Gamma(r+x)}{x!\Gamma(r)} (1-p)^r p^x,
$$
for parameters $r>0$ and $p\in(0,1)$. We write $X\sim \text{NB}(r,p)$. For positive integer $r$, the negative binomial distribution can be recognized as the distribution for the number of heads before the $r$th tail in biased coin-tossing, but it is a valid distribution for all $r>0$. In engineering, it is sometimes called the P\'olya distribution in the case where $r$ is not integer.

\subsection{Derivation of the Expectation/Conditional Maximixation (ECM) Algorithm in Section \ref{subsec:ECM}}
Treating $\lambda = \lambda_{iq}$ for all $i$ and $q=2$ to $Q$ as missing data and $y = y_{ijq}$ for all $i$, $j$ and $q$ as observed data, the complete data is $(y_{ijq},\lambda)$. Denote $\theta = (\gamma,(\beta_q)_{q=2}^Q)$, where $\gamma$ includes the incidental parameters $\log(\delta_{ij})$ for all $i$ and $j$. The complete data log-likeliood $\ell(\theta|y,\lambda)$ is
\begin{gather} 
-\sum_i\sum_{q\neq1} \lambda_{iq} \left( \sum_j e^{x_{ijq}^T\gamma} \right) + \sum_i\sum_{q\neq1} y_{i+q}\log\lambda_{iq} + \sum_i\sum_j\sum_q x_{ijq}^T\gamma y_{ijq} + \sum_i\sum_j\sum_q \log(y_{ijq}!) +  \nonumber \\
\sum_i\sum_{q\neq1} (1/\beta_q-1)\log\lambda_{iq} - \sum_i\sum_{q\neq1} \lambda_{iq}/\beta_q - \sum_i\sum_{q\neq1} \log\beta_q/\beta_q-\sum_i\sum_{q\neq1} \log\Gamma(1/\beta_q). 
\label{eqn:loglik_completedata}
\end{gather} \

The $(t+1)$th E-step involves finding the conditional expectation of the complete data log-likelihood with respect to to the conditional distribution of $\lambda$ given $y$ and the current estimated parameter $\theta^{(t)}$. Straightforward algebra establishes that
\begin{equation}
\lambda_{iq}|y_{ijq},\theta^{(t)} \sim \mathcal{G} \left( y_{i+q}+1/\beta_q^{(t)} , \left(\sum_j e^{x_{ijq}\gamma^{(t)}}+1/\beta_q^{(t)}\right)^{-1} \right),
\end{equation}
\noindent independently for each $i$ and $q$, where the gamma distribution is parameterized in terms of scale parameter. It follows that
\begin{gather}
\hat{\lambda}_{iq}^{(t+1)} = \text{E}\left(\lambda_{iq}\middle|(y_{ijq})_j,\theta^{(t)}\right) = \frac{y_{i+q}+1/\beta_q^{(t)}}{\sum_j e^{x_{ijq}\gamma^{(t)}}+1/\beta_q^{(t)}} \\
\hat{\chi}_{iq}^{(t+1)} = \text{E}\left(\log(\lambda_{iq})\middle|(y_{ijq})_j,\theta^{(t)}\right) = \psi\left(y_{i+q}+1/\beta_q^{(t)}\right) - \log\left(\sum_j e^{x_{ijq}\gamma^{(t)}}+1/\beta_q^{(t)}\right).
\end{gather} \

Thus, in the $(t+1)$th E-step, we replace $\lambda_{iq}$ and $\chi_{iq} = \log(\lambda_{iq})$ in Equation \ref{eqn:loglik_completedata} with $\hat{\lambda}_{iq}^{(t+1)}$ and $\hat{\chi}_{iq}^{(t+1)}$, giving $Q(\theta|\theta^{(t)})$. The $(t+1)$th CM-step then finds $\theta^{(t+1)}$ to maximize $Q(\theta|\theta^{(t)})$ via a sequence of conditional maximization steps, each of which maximizes the $Q$ function over a subset of $\theta$, with the rest fixed at its previous value. In our application, it is natural to partition $\theta$ into $\gamma$ and $\beta_q$ for each $q=2$ to $Q$. Differentiating Equation \ref{eqn:loglik_completedata} with respect to $\gamma$, we obtain
\begin{equation}
-\sum_i\sum_j\sum_q \lambda_{iq} x_{ijq} e^{x_{ijq}^T\gamma} + \sum_i\sum_j\sum_q x_{ijq} y_{ijq},
\end{equation}

\noindent which is the score equation of the Poisson log-linear model \cite[]{Mccullagh1989generalized} with an additional offset $\lambda_{iq}$. This allows us to leverage existing functions for fitting generalized linear models available in most statistical software packages for maximizing $\gamma$ in the CM step. This is an important feature as $\gamma$ often contains a huge amount of parameters in our applications, due to the inclusion the incidental parameter $\log(\delta_{ij})$ for every unique combination of covariates. Existing functions for fitting generalized linear models are typically stable and heavily optimized, even for a large number of parameters. Maximizing $\beta_q$ in the CM step for each q is straightforward, as it only involves univariate optimization.

\section*{Acknowledgements}
Lee's research is partially supported by the Australian Bureau of Statistics. The authors are grateful to Zhen Chen for providing the yogurt data.


\newpage

\bibliographystyle{agsm} 

\bibliography{../phd} 

\end{document}